\documentclass[lettersize,journal]{IEEEtran}

\usepackage{amsmath}
\usepackage{amsfonts}
\usepackage{amssymb}
\usepackage{amsthm}
\usepackage{tabularx}
\usepackage{mathrsfs} 
\usepackage{soul}
\usepackage{graphicx}

\usepackage{url}
\usepackage{hyperref}

\usepackage{stfloats}
\usepackage{float}
\usepackage{graphicx}
\usepackage{cite}
\usepackage{xcolor}
\usepackage{subfigure}

\usepackage{cleveref}

\usepackage{threeparttable}

\usepackage{algorithm}
\usepackage{algpseudocode}

\usepackage{pifont}

\usepackage{booktabs}

\usepackage{multirow}

\usepackage{rotating}

\usepackage{makecell}

\usepackage{array}

\def\BibTeX{{\rm B\kern-.05em{\sc i\kern-.025em b}\kern-.08em
    T\kern-.1667em\lower.7ex\hbox{E}\kern-.125emX}}
\usepackage{balance}

\newtheorem{remark}{Remark}

\begin{document}
\title{Voltage Profile-Driven Physical Layer Authentication for RIS-aided Backscattering Tag-to-Tag Networks}
\author{Masoud~Kaveh\IEEEmembership{}, 
Farshad Rostami Ghadi, \IEEEmembership{Member, IEEE}, Yifan Zhang, \IEEEmembership{Graduate Student Member,~IEEE,}  \\
Zheng Yan, \IEEEmembership{Fellow, IEEE}, and  Riku Jäntti, \IEEEmembership{Senior Member, IEEE}
\thanks{This work is supported in part by the Academy of Finland under Grants 345072 and 350464.

M. Kaveh, Y. F. Zhang, and R. Jäntti are with the Department of Information and Communication Engineering, Aalto University, Espoo, Finland. (e-mail: $\rm masoud.kaveh@aalto.fi, yifan.1.zhang@aalto.fi, riku.jantti@aalto.fi$)
	 	
F. R. Ghadi is with the Department of Electronic and Electrical Engineering, University College London, WC1E 6BT London, UK. (e-mail: $\rm f.rostamighadi@ucl.ac.uk$).

Z. Yan is with the School of Cyber Engineering, Xidian University, 710071 Xi'an, China, (e-mail: $\rm zyan@xidian.edu.cn$)}}


\maketitle

\begin{abstract}
Backscattering tag-to-tag networks (BTTNs) are emerging passive radio frequency identification (RFID) systems that facilitate direct communication between tags using an external RF field and play a pivotal role in ubiquitous Internet of Things (IoT) applications. 
Despite their potential, BTTNs face significant security vulnerabilities, which remain their primary concern to enable reliable communication. 
Existing authentication schemes in backscatter communication (BC) systems, which mainly focus on tag-to-reader or reader-to-tag scenarios, are unsuitable for BTTNs due to the ultra-low power constraints and limited computational capabilities of the tags, leaving the challenge of secure tag-to-tag authentication largely unexplored.
To bridge this gap, this paper proposes a physical layer authentication (PLA) scheme, where a Talker tag (TT) and a Listener tag (LT) can authenticate each other in the presence of an adversary, only leveraging the unique output voltage profile of the energy harvesting and the envelope detector circuits embedded in their power and demodulation units. This allows for efficient authentication of BTTN tags without additional computational overhead. 
In addition, since the low spectral efficiency and limited coverage range in BTTNs hinder PLA performance, we propose integrating an indoor reconfigurable intelligent surface (RIS) into the system to enhance authentication accuracy and enable successful authentication over longer distances.
Security analysis and simulation results indicate that our scheme is robust against various attack vectors 
and achieves acceptable performance across various experimental settings. Additionally, the results indicate that using RIS significantly enhances PLA performance in terms of accuracy and robustness, especially at longer distances compared to traditional BTTN scenarios without RIS.

\end{abstract}


\section{Introduction}
\IEEEPARstart{I}{n} an era of rapidly expanding connectivity, the demand for energy-efficient communication systems and zero-energy devices continues to rise, particularly for applications where regular battery replacement or recharging is impractical \cite{zero6G1}. Radio frequency identification (RFID) technology has become a prime example of this advancement, as it allows the identification and tracking of tags using electromagnetic fields, all without the need for an internal power source \cite{RFID-ManKind}. These tags draw energy from the reader’s transmitted signals and respond by backscattering a modulated signal containing the relevant data \cite{BCFAS1}.
A breakthrough in this field is the development of backscatter communication (BC), which enables devices to communicate by reflecting existing RF signals rather than generating new transmissions. This approach significantly lowers power consumption, making it well-suited for passive RFID systems \cite{BCRFID1,BCArbit1}. Nevertheless, despite the benefits of RFID and BC systems, they face notable limitations, particularly in enabling direct communication between individual tags.

In traditional RFID and BC setups, communication typically occurs between a reader and a tag, which hinders scalability and flexibility in certain scenarios \cite{BCSI1}. For example, many applications, such as large-scale inventory management and smart environments, require tags to exchange information directly without needing constant intervention from a central reader. This has led to the development of backscattering tag-to-tag networks (BTTNs), which enable direct communication between tags within an external RF field \cite{BTTN-design-1}.
BTTNs represent a novel approach in passive RFID systems, facilitating communication between tags such as a \textit{talker tag} (TT) and a \textit{listener tag} (LT), with the help of an external RF source \cite{design-BTTN-2}. In this system, TT modulates the RF field by adjusting its load impedance, thus altering the characteristics of the backscattered signal directed at LT. The LT then demodulates this backscattered signal using its envelope detector. Additionally, both tags incorporate energy harvesting circuits that convert RF energy into electrical power, ensuring continuous operation without the need for external power sources \cite{design-BTTN-3}.

However, BTTNs face several significant challenges that must be overcome to ensure efficient and reliable operation. One of the primary challenges is the requirement for ultra-low power consumption. Since BTTNs rely on only passive tags at both communication endpoints that harvest energy from incident RF signals, it is crucial to minimize the power consumption to extend the operational lifetime of the tags and maintain their functionality.
Another challenge is the low spectral efficiency of BC. The process of modulating and reflecting the RF signal inherently limits the bandwidth available for communication, resulting in lower data rates compared to active transmission methods. This issue is exacerbated in BTTNs compared to BC and RFID systems, as the receiver is not a powerful reader but another tag with comparatively fewer processing resources.
Furthermore, BTTNs are constrained by very short communication ranges. The passive nature of the tags and the reliance on backscattered signals result in a limited communication distance, which can be a significant drawback in applications requiring long-range interactions. 
Last but not least, since current BTTNs are inherently vulnerable to security threats, it is crucial that tags authenticate each other to ensure secure communication.

\subsection{Related Works} \label{sec-RelWo}

The exploration of PLA has emerged as an effective alternative to cryptographic methods for authentication in BC by utilizing the unique physical properties of backscattered signals. Over recent years, diverse strategies have been proposed to enhance the security of BC systems by leveraging PLA.
In \cite{BC_PLA_01}, a PLA technique was introduced to identify UHF RFID tags, focusing on security and reliability in BC systems by exploiting the unique inherent physical properties of each tag. GenePrint \cite{BC_PLA_02} proposed a versatile PLA framework applicable to various RFID systems, with the aim of improving security and operational efficiency. Similarly, \cite{BC_PLA_03} addressed identity attack vulnerabilities in BC systems by proposing a PLA mechanism to fortify the system security and maintain the integrity of the data.
Wang et al. \cite{BC_PLA_04} introduced a replay-resistant authentication approach that combined signal randomization and tag coupling to counteract replay, signal compensation, and brute-force attacks. Danev et al. \cite{BC_PLA_05} conducted an extensive study on PLA of RFID devices, presenting multiple methods to derive physical layer fingerprints and demonstrating high accuracy in identifying RFID transponders according to their physical characteristics. To address active attack vulnerabilities in BDs, the authors of \cite{BC_PLA_06} developed ShieldScatter, a low-cost system that generates unique multipath propagation signatures, allowing sensitive source identification and threat detection.
BCAuth \cite{BCAuth}, proposed by Wang et al., implemented a multistage authentication and attack tracing scheme. It utilized the spatial information of BDs for initial and re-authentication processes, enhancing security for both static and mobile BDs. Li et al. \cite{PLA_AmBC_NOMA} explored PLA for ambient BC-aided NOMA systems, introducing three schemes based on multiplexing forms: shared authentication tags, space-division multiplexing tags, and time-division multiplexing tags. The study also examined detection probabilities and false alarm rates under channel estimation errors.
Yang et al. \cite{BatAu} developed BatAu, a batch authentication framework for smart home networks, leveraging physical-layer attributes in multiplexed signals to authenticate multiple BDs simultaneously. Zhang et al. \cite{Cross-Domain-PLA} introduced FedScatter, a lightweight cross-domain PLA system for BC environments. FedScatter constructed device identity signatures from passive signal attributes and used federated learning to aggregate cross-domain identity data while preserving user privacy.
In \cite{APAuth}, the authors proposed APAuth, a lightweight authentication scheme designed for access points using backscatter devices. The approach takes advantage of the unique characteristics of the power volumes harvested from the tags to authenticate access points effectively without the need for complex computations.
In \cite{BatchAuth}, the authors introduced BatchAuth, a PLA scheme for multiple tags in smart environments. This scheme utilizes joint CSI and physical layer features to authenticate a batch of BDs simultaneously, to effectively counter impersonation and replay attacks while adapting to the mobility of devices in dynamic environments.

\begin{table} [t]
    \centering
    \begin{threeparttable}
    \caption{Comparison of Our Work with Existing PLA Schemes for BC Systems} 
    \label{table_1}
    {
    \begin{tabular}{c|ccccccc}
    \hline \hline
      Works   & AT & TAT & ESCR & CFA & MA & CSAI & PLA \\
    \hline
    
    \cite{BC_PLA_04, BC_PLA_05, BC_PLA_06} & $\times$ & $\times$ & $\times$ & \checkmark & $\times$ & $\times$ & \checkmark\\
    \hline
        \cite{BCAuth, Cross-Domain-PLA} & $\times$ & $\times$ & $\times$ & $\otimes$ & $\otimes$ & $\times$ & \checkmark \\
    \hline
        \cite{PLA_AmBC_NOMA, BatAu, BatchAuth} & $\times$ & $\times$ & $\times$ & \checkmark & $\times$ & $\otimes$ & \checkmark \\
        \hline
        \cite{APAuth} & \checkmark & $\times$ & $\times$  & $\otimes$ & $\times$ & $\otimes$ & $\otimes$  \\          \hline    
        Ours & \checkmark & \checkmark & \checkmark & \checkmark & \checkmark & \checkmark & \checkmark \\
    \hline
    \end{tabular}
    } 
    \begin{tablenotes}
        \item AT: authentication at tag , TAT: tag authenticated by another tag, ESCR: expanded secure coverage range, CFA: cryptography-free approach, MA: mutual authentication, CSAI: complex signal analyzer independence, PLA: physical layer attributes  \checkmark: Item is supported, $\times$: Item is not supported, $\otimes$: Item is conditionally supported. 
    \end{tablenotes}
\end{threeparttable}
\end{table}

\subsection{Research Gaps and Motivations}

Although various authentication schemes have been proposed for BC and FRID systems, where a reader authenticates a tag or vice versa, the challenge of enabling tag-to-tag authentication remains unexplored.
Since tags in BTTNs are ultra-low-power devices designed to operate with minimal energy consumption, they often lack the computational capacity to process lightweight cryptographic primitives or even compute physical layer attributes such as received signal strength (RSS) and angle of arrival (AoA). This limitation severely restricts the ability to implement the same traditional PLA frameworks used in BC for BTTNs.
Furthermore, the combination of low spectral efficiency and limited processing capability keeps the authentication performance even lower in BTTNs compared to typical BC and RFID systems.
The short coverage range of BTTNs compounds these challenges; as the distance between tags increases, the quality of the received signal decreases, making it even more difficult to achieve reliable authentication over longer distances. 

These combined factors: security vulnerabilities, ultralow power constraints, low spectral efficiency, and short communication range make it extremely challenging to conduct secure communication in BTTNs. To date, no effective authentication scheme for BTTNs has been proposed in the literature, largely due to these challenges. Our research aims to address this gap by proposing a lightweight and practical physical layer authentication (PLA) scheme for BTTNs.
The proposed PLA scheme takes advantage of the unique behavior of the signals in the internal circuits of the tags to enable efficient and secure authentication without imposing additional computational burdens. Additionally, to enhance the quality of the received signal power at LT and, thereby, improve the performance of the PLA, we introduce integrating an indoor reconfigurable intelligent surface (RIS) into BTTN. Incorporating RIS helps to improve signal quality and extend the effective communication range, ensuring that LT can authenticate TT over longer distances.
Table \ref{table_1} presents the unique aspects of our article compared to related works.

\subsection{Contributions}


This paper proposes a novel PLA scheme to overcome the limitations of existing authentication methods in BC to enable secure communication in BTTNs. 
Using the unique voltage profiles generated by the energy harvesting and demodulator circuits on the tags, our scheme provides a lightweight but robust solution to make the tags authenticate each other without imposing additional computational overhead. Furthermore, we show that the integration of a RIS into the system enhances authentication accuracy and extends the secure communication range, addressing the inherent PLA performance constraints within BTTNs.
The key contribution of this paper can be presented as follows. 

\begin{itemize}
    \item In this paper, we propose the first authentication scheme for BTTNs. Our approach leverages the unique output voltage profile generated by the energy harvesting and envelope detector circuits in the power and demodulation units embedded in LT. Using existing measurements within these circuits, our scheme creates a distinct signature for each tag, enabling a straightforward and efficient authentication of TT without imposing any additional computational overhead.
    
    \item To improve the reliability of the created voltage profiles in TT, enhance RSS, and boost the performance of the proposed PLA scheme, we introduce the use of indoor RIS in BTTN communication system. The addition of RIS can also extend the range of secure communication within BTTNs.
    
    \item We perform a comprehensive security analysis under different threat scenarios, showing that the proposed PLA allows LT to effectively authenticate TT in the presence of an adversary. Furthermore, we show that incorporating RIS into BTTN enhances the security of the authentication process by making it challenging for the adversary to replicate the same output voltage profile at LT.
    
    \item The proposed authentication scheme is evaluated through extensive experiments in various system configurations. The results confirm that our scheme achieves acceptable authentication performance across different settings. Additionally, the findings indicate that the inclusion of RIS significantly enhances the authentication performance, particularly over longer distances, compared to BTTN scenarios without RIS.
\end{itemize}

\subsection{Organization}
The remainder of this paper is organized as follows. 
Section \ref{sec_BTTN} provides an overview of BTTNs and the key components within the tags. 
Section \ref{sec_sys} presents the system and threat models. In Section \ref{sec_RISAuth}, we elaborate on the proposed authentication scheme. Section \ref{sec_analy} provides a comprehensive security analysis of the authentication scheme, examining its robustness against various attack vectors. The simulation results are presented in Section \ref{sec_results}, and finally, Section \ref{sec_conclusion} concludes the paper.

\section{Backscattering Tag-to-Tag Networks} \label{sec_BTTN}

This section provides a review of the BTTN architecture and discusses the critical components within the BTTN tags that enable efficient and passive communication.

\begin{figure}[t]
    \centering    \includegraphics[width=0.31\textwidth]{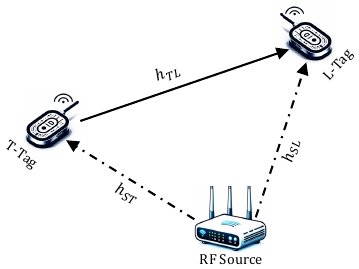}
    \caption{A BTTN communication system where a TT tries to talk to an LT by modulating and backscattering the signal from the RF source.}
    \label{fig:BTTNsysmodel}
\end{figure}
\subsection{BTTN Overview}
The primary components of a BTTN typically include a TT, LT, and an external RF source, as illustrated in Fig. \ref{fig:BTTNsysmodel}. This system leverages the existing RF environment to facilitate communication between tags by modulating and backscattering RF signals.
In a typical BTTN scenario, TT modulates its impedance, which changes the reflection coefficient of the tag antenna, thereby encoding information in the backscattered signal. Then LT receives this modulated backscattered signal \cite{BTTN-performance-1,passive-BTTN-1}.
The signal received at LT is a combination of the unmodulated signal from the RF source and the backscattered signal from TT. The signal received at LT can be expressed as:
\begin{align}
Y_\text{L} = \mathcal{B}_\text{T} \sqrt{P_\text{s}} \left( h_\text{ST} h_\text{TL} + h_\text{SL} \right) +n_\text{L} ,
\end{align}
where \( \mathcal{B}_\text{T} \) is the backscattering coefficient of TT as a function of reflection
coefficient of the antenna at different modulation states,
     \( P_\text{s} \) is the power of the signal emitted by the RF source,
     \( h_\text{ST} \) is the channel coefficient between the RF source and TT,
    \( h_\text{TL} \) is the channel coefficient between TT and LT,
    \( h_\text{SL} \) is the channel coefficient between the RF source and the LT, and $n_\text{L}$ shows the additive white Gaussian noise (AWGN) at LT with zero mean and variance \( \sigma^2_\text{T} \).
This received signal is then processed by LT's demodulator circuit to extract the information sent by TT.
If LT needs to communicate with TT, the process is similar, but in reverse. LT modulates its impedance to backscatter the RF signal, and TT receives this modulated backscattered signal. Then, the received signal at TT can be expressed as:
\begin{align}
Y_\text{T} = \mathcal{B}_\text{L} \sqrt{P_\text{s}} \left( h_\text{SL} h_\text{TL} + h_\text{ST} \right) +n_\text{T} ,
\end{align}
where \( \mathcal{B}_\text{L} \) is the backscattering coefficient of LT and $n_\text{T}$ shows the noise at TT with zero mean and variance \( \sigma^2_\text{T} \).

\subsection{BTTN Tag Design}
This section provides an overview of the fundamental components of a passive BTTN tag, which operates without an active onboard transmitter. 
\begin{figure}[t]
    \centering    \includegraphics[width=0.42\textwidth]{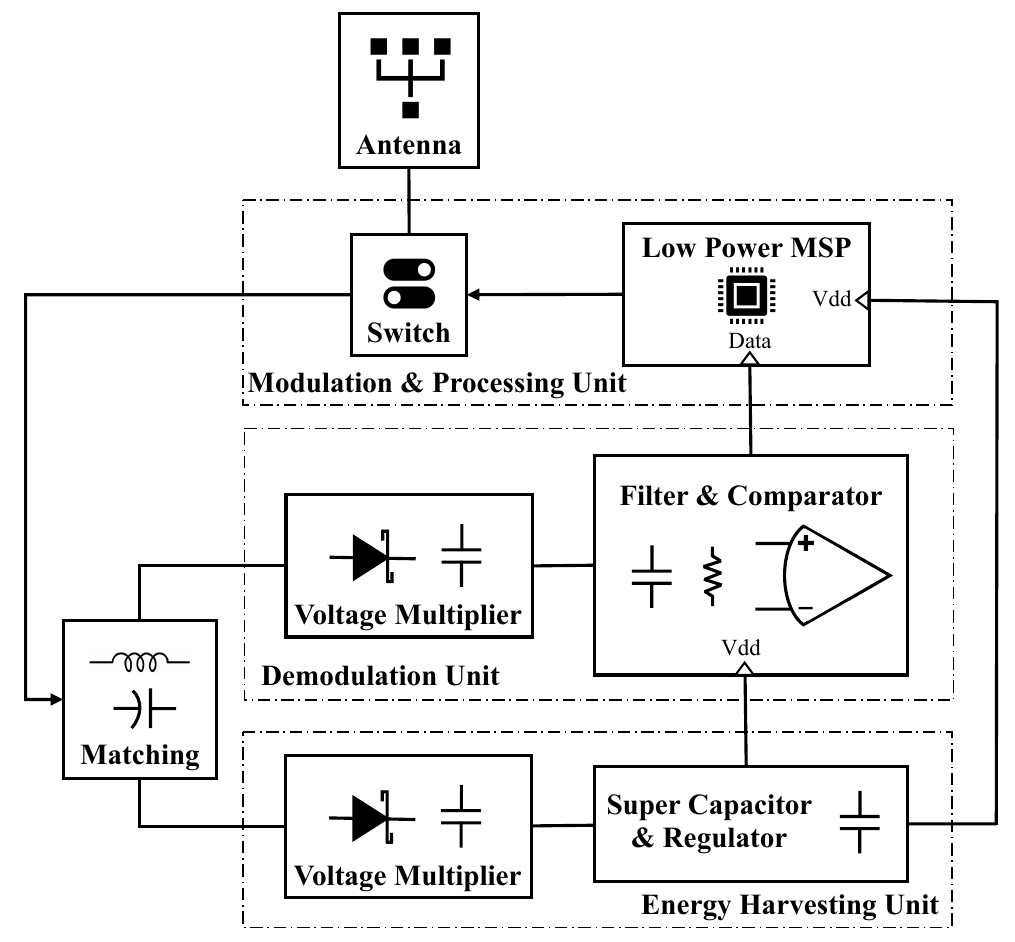}
    \caption{A basic circuit block diagram of a tag in BTTN.}
    \label{fig:Tag_Cir_Blck}
\end{figure}
As depicted in Fig. \ref{fig:Tag_Cir_Blck}, a basic BTTN tag comprises an antenna, a power harvesting circuit, an envelope detector (serving as a zero-consumption receiver), a modulator (acting as a low-power passive transmitter), and a low power MSP microcontroller responsible for data processing.
When a tag functions as a transmitter (TT), the modulator adjusts the excitation field by altering the load impedance \( Z_\text{T} \) of the antenna to backscatter signals. Conversely, when a tag operates as a receiver (LT), the envelope detector demodulates the received signal by detecting the voltage difference between the backscattered signals at distinct modulation states. This behavior can be characterized by the following differential radar cross-section (RCS) equation:
\begin{align}
    \Delta \sigma = \frac{\lambda^2 G_\mathrm{tag}^2}{4 \pi} \left| \Gamma^*_\mathrm{L,1} - \Gamma^*_\mathrm{L,2} \right|^2,
\end{align}
where \( \lambda \) is the wavelength of the excitation signal, $ G_\mathrm{tag}$ is the antenna gain of the tag, and \( \Gamma^*_\mathrm{L,i} \) is the conjugate reflection coefficient of the antenna at different modulation states. 
The reflection coefficient is described as:
\begin{align}   
\Gamma^*_\mathrm{L,i} = \frac{Z_\mathrm{L,i} - Z^*_\text{a}}{Z_\mathrm{L,i} + Z^*_\text{a}}, \quad \mathrm{i = on, off},
\end{align}
where \( Z_a \) denotes the antenna impedance. The modulator typically consists of a switch that connects to various terminating impedances, depending on the digital modulation used.
\begin{figure}[h]
    \centering    \includegraphics[width=0.26\textwidth]{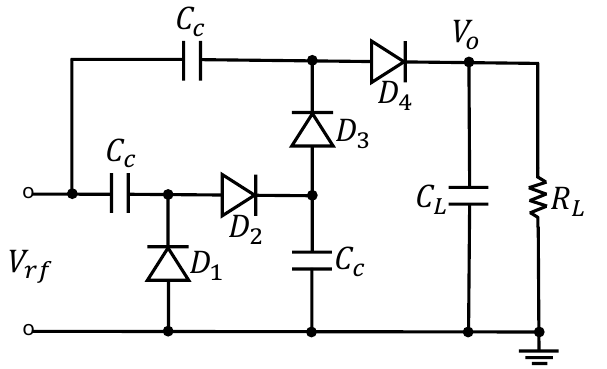}
    \caption{Two-stage voltage multiplication circuit for energy harvesting.}
    \label{fig:EH_Circuit}
\end{figure}
The power harvesting circuit, illustrated in Fig. \ref{fig:EH_Circuit}, includes a voltage multiplier that converts a low input voltage into a higher, regulated voltage suitable for powering the demodulation and decoding circuits. To optimize power transfer from the antenna and minimize reflected power, the voltage multiplier is preceded by an impedance matching network. However, the input impedance of the power harvesting circuit is only matched to the antenna at a specific input power due to its dependence on the received power level. Therefore, the impedance is tuned to match the minimum input power required for operation. At higher input power levels, the mismatch causes some power to be reflected, reducing overall power efficiency. A voltage regulator follows the voltage multiplier to provide a stable supply voltage for the demodulation and decoding logic circuits.
\begin{figure}[h]
    \centering    \includegraphics[width=0.41\textwidth]{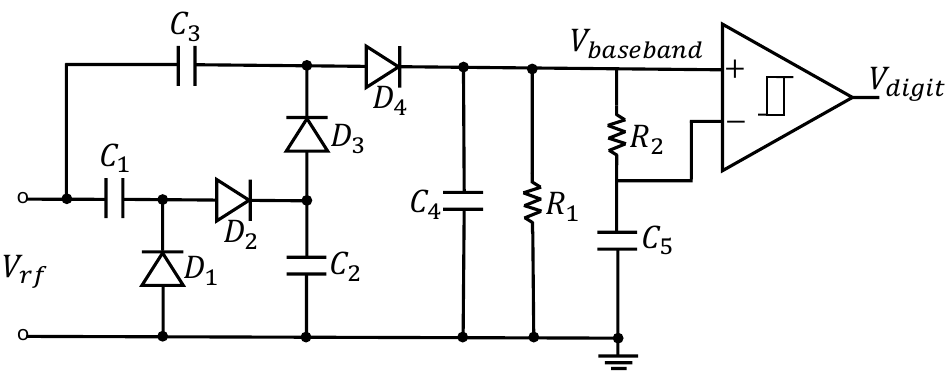}
    \caption{Demodulator circuit in a BTTN tag.}
    \label{fig:Dem_Circuit}
\end{figure}
The analog demodulation section, shown in Fig. \ref{fig:Dem_Circuit}, consists of a voltage multiplier followed by an impedance matching circuit and an envelope detector, which extracts the baseband signal. The subsequent circuitry distinguishes between different voltage levels in the baseband signal that correspond to variations in the input power at the antenna. This difference is then digitized by a comparator, typically acting as a data slicer, comparing the baseband signal with its own average for further processing in the digital platform.

The co-design of the energy harvesting and demodulation circuits is crucial due to their interdependence. Since both circuits draw power from the same source, the allocation of power to each circuit directly impacts their performance. Therefore, finding the optimal division of power between these two circuits is essential for enhancing the communication range between the tags.
For the energy harvesting circuit, the selection of Schottky diodes and the number of stages in the voltage multiplier are key design choices that affect efficiency. While more stages typically yield higher voltages, it has been demonstrated that in low power environments, fewer stages can be more efficient than having a larger number of stages \cite{ref-tag-EH-design}. To ensure sufficient power for tag operation without requiring a storage element, it has been shown that the input power must be approximately $-15$ dBm \cite{ref-tag-EH-design}.
In the case of the demodulation circuit, determining the appropriate modulation depth in BTTN is essential. To address this, a link budget analysis can be performed, assuming that the excitation signal is evenly distributed and identical throughout the environment. For simplicity, it is assumed that the input power \( P_{\text{tag}} \) received by the tags is constant across the environment and TT is in phase with the excitation signal \cite{ref-tag-dem-phase}.

\section{System and Threat Models} \label{sec_sys}
In this section, we explain the system, channel, and threat models to determine the security goals for the studied RIS-aided BTTN.

\subsection{System and Channel Model}
\begin{figure}[t]
    \centering    \includegraphics[width=0.40\textwidth]{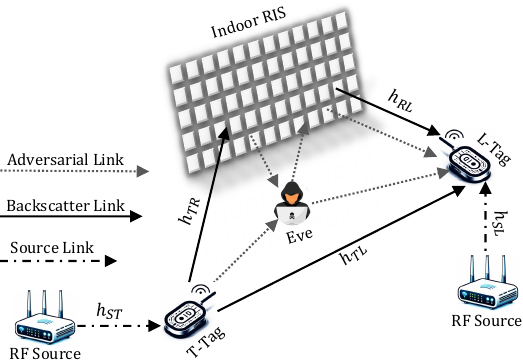}
    \caption{System model and security model depicting the tag, reader, and RIS configuration in RIS-aided MBC.}
    \label{fig:sysmodel}
\end{figure}

We consider a BTTN setup that includes two tags (TT and LT) and their corresponding RF sources, an attacker (Eve), and a RIS, as depicted in Fig.~\ref{fig:sysmodel}. 
In this configuration, both TT and LT serve as semi-passive backscatter devices, powered by continuous wave carrier signals emitted by RF sources in an indoor environment. The primary goal of TT is to backscatter its information to LT, with the assistance of an indoor RIS composed of \emph{N} reflective elements. In traditional BTTN operations, TT modifies its load impedance, influencing the reflection coefficient of its antenna to encode data within the backscattered signal, which is subsequently detected by LT \cite{BTTN-CSI-1}. The signal received by LT consists of both the direct unmodulated signal from the RF source and the modulated backscattered signal from TT, which travels through both direct and RIS-assisted paths.
Without loss of generality, it is assumed that the RF source transmits unmodulated carrier signals. This allows both tags to apply simple cancellation techniques to mitigate interference originating from the source link \cite{Riku_DirPathInterf, Riku_IEEERFID1}. It is further assumed that the RIS has knowledge of the channel state information (CSI) for the cascaded link and the microcontroller-enabled RIS operates under a near optimal condition, enabling the optimization of the phase shifts of its reflective elements to maximize the signal-to-noise ratio (SNR) at the receiver's end \cite{BjornsonCSI, RIS-PLS-SG}. 
For simplicity, TT and LT are considered to be equipped with a single antenna each. Therefore, the signal received at TT is expressed as follows:
\begin{align}
Y_\text{T}=\sqrt{P_\text{s}}h_\text{ST}+n_\text{T},
\end{align}
Since the noise caused by the tag’s antenna is significantly lower than the received signal from the source, it is neglected for the remainder of this paper \cite{ref57}.
The backscattered signal received by LT can be written as:
\begin{align} \label{eq-LT-sig}
Y_\text{L} = \mathcal{B}_\text{T} S(t) \sqrt{P_\text{s}} \left( h_\text{ST} h_\text{TL} + h_\text{ST} \mathbf{H}_\mathrm{RL}^{\mathcal{T}} \mathbf{\Phi}  \mathbf{H}_\mathrm{TR} \right) +n_\text{L} ,
\end{align}
where \( S(t) \) represents the unit power information signal backscattered by TT. \( \mathbf{H}^{\mathcal{T}} \) indicates the transpose of matrix \( \mathbf{H} \), with \( \mathbf{H}_\mathrm{TR} \) being the channel coefficient between TT and RIS, and \( \mathbf{H}_\mathrm{RL} \) representing the channel coefficient between RIS and LT. The matrix \( \mathbf{\Phi} \) refers to the phase shift matrix of RIS, which can be defined as \( \mathbf{\Phi}=\text{diag}\left(\left[\mathrm{e}^{j\phi_1}, \mathrm{e}^{j\phi_2},...,\mathrm{e}^{j\phi_N}\right]\right) \), where \( N \) is the number of RIS elements.
The RIS-aided channel coefficients, \( \mathbf{H}_\mathrm{TR} \) and \( \mathbf{H}_\mathrm{RL} \), represent the \( N \) coefficients from TT to RIS and RIS to LT, respectively. These can be expressed as \( \mathbf{H}_\mathrm{TR}=d_\mathrm{TR}^{-\chi}.\left[h_\mathrm{TR_1}\mathrm{e}^{-j\delta_1}, h_\mathrm{TR_2}\mathrm{e}^{-j\delta_2},..., h_\mathrm{TR_N}\mathrm{e}^{-j\delta_N}\right] \) and \( \mathbf{H}_\mathrm{RL}=d_\mathrm{RL}^{-\chi}.\left[h_\mathrm{RL_1}\mathrm{e}^{-j\zeta_{1}}, h_\mathrm{RL_2}\mathrm{e}^{-j\zeta_{2}},..., h_\mathrm{RL_N}\mathrm{e}^{-j\zeta_{N}}\right] \), where \( j=\sqrt{-1} \) and \( d_\mathrm{TR} \) and \( d_\mathrm{RL} \) are the distances between TT and RIS, and RIS and LT, respectively \cite{V2V-RIS-TIV}. \( \chi \) is the path-loss exponent, \( h_\mathrm{TR_n} \) and \( h_\mathrm{RL_n} \) are the channel coefficient amplitudes, and \( \mathrm{e}^{-j\delta_n} \) and \( \mathrm{e}^{-j\zeta_{n}} \) are the phase shifts for each RIS element, where \( n\in\left\{1,2,...,N\right\} \).
Given the short transmission range nature of BC in BTTNs, it is assumed that all links experience Rician fading. The backscattered signal is then received by the energy and envelope detectors' circuits LT to authenticate and retrieve the information transmitted by TT.

\subsection{Threat Model} \label{subsec:threat}
Given the absence of cryptographic processing onboard and the open nature of wireless channels, BTTNs are highly susceptible to various security threats. Therefore, security policies must ensure the confidentiality, integrity and authenticity of the communication between TT and LT while preventing adversarial access to RIS.
It is important to note that indoor RISs are typically designed with BC scenarios in mind, making it difficult for adversaries to physically manipulate the microcontroller \cite{Indoor_RIS_01, Indoor_RIS_02}. As a result, our PLA design operates under the assumption that the RIS is either regarded a trusted component or is managed by a reliable entity.
The primary security goal in BTTNs is to ensure that only legitimate tags (LT and TT) are able to exchange information securely and that any attempts by an adversary (Eve) to interfere, intercept, or impersonate a legitimate tag are detected and thwarted. Eve is regarded as a highly capable adversary with access to the communication channels, enabling her to launch the following common attack strategies against BTTNs:

\subsubsection{Impersonation Attacks}
In an impersonation attack, Eve attempts to act as the TT by transmitting signals that closely mimic the backscattered signals of the legitimate TT. If Eve can generate a backscattered signal similar enough to the legitimate TT’s profile signal, LT might fail to distinguish Eve from TT, enabling Eve to pass the authentication test and leading to a successful impersonation attack. 

\subsubsection{Man-in-the-Middle (MITM) Attacks}
An MITM attack involves Eve positioning herself between TT and LT, intercepting and possibly modifying the communications between the two tags;
making it appear as if the altered message came directly from TT. Since BTTNs rely on passive communication and signal backscattering, an MITM attack could exploit the weak and easily intercepted nature of the signals. 

\subsubsection{Replay Attack}
In a replay attack, Eve captures a valid transmission from TT to LT and replays it later to deceive LT into believing that the message came from the legitimate TT. Since the replayed signal is identical to the original one, LT could be fooled into authenticating Eve’s transmission as genuine. In the context of BTTNs, replay attacks are especially dangerous because the signals are usually simplistic and lack dynamic cryptographic signatures or timestamps. 

\subsubsection{Relay Attacks}
Relay attacks involve Eve acting as an intermediary between TT and LT, relaying signals from TT to LT from a different location. In this attack, Eve may either forward the original message from TT at the same time or block it entirely, then relay it to deceive LT. 
This type of attack exploits the passive nature of BC, as Eve can artificially extend the communication range by relaying messages from a location closer to LT, potentially making the transmission appear legitimate.


The primary security goal of this paper is to
ensure that LT can reliably authenticate the legitimacy of TT (and optionally, vice versa) while preventing impersonation attempts by an adversary.
    Moreover, the proposed PLA aims to guarantee that the messages exchanged between TT and LT are not tampered with during transmission, protect the content of the messages from being intercepted or overheard by unauthorized parties, and ensure that previously transmitted or relayed signals cannot be reused to gain unauthorized access.

\section{The Proposed PLA for RIS-aided BTTNs} \label{sec_RISAuth}
This section elaborates on the proposed RIS-aided PLA design for BTTNs, including output voltage derivation to approximate the measured voltage profiles at tags, initialization phase, and authentication phase.

\subsection{Output Voltage Derivation}
Since the proposed authentication scheme relies on the voltage profile signature at LT, this section provides detailed insights, focusing on deriving the output voltages from the energy harvesting and demodulator circuits in RIS-aided BTTNs to establish an accurate representation of the measured voltage profiles.
According to \eqref{eq-LT-sig}, the RSS at LT can be determined as
\newcommand\myeq{\mathrel{\stackrel{\makebox[0pt]{\mbox{\normalfont\small def}}}{=}}}
\newcommand\myapprox{\mathrel{\stackrel{\makebox[0pt]{\mbox{\normalfont\footnotesize (a)}}}{\approx}}}
\begin{align} 
&\gamma_L={\left|\sqrt{P_\text{s}} \left( h_\text{ST} h_\text{TL} + h_\text{ST} \mathbf{H}_\mathrm{RL}^{\mathcal{T}} \mathbf{\Phi}  \mathbf{H}_\mathrm{TR} \right)\right|}^2  \\ 
&\hspace{-0.2cm}\approx\frac{P_\text{s}|h_\mathrm{ST}|^2|h_\mathrm{TL}|^2}{{d}^\chi_\mathrm{ST}{d}^\chi_\mathrm{TL}}
\hspace{-2pt} +\hspace{-2pt}
\frac{P_\mathrm{s}|h_\mathrm{ST}|^2\left|\sum_{n=1}^{N}h_\mathrm{TR_n} h_\mathrm{RL_n} \mathrm{e}^{j\Psi_n}\right|^2}{{d}^\chi_\mathrm{ST}{d}^\chi_\mathrm{TR}{d}^\chi_\mathrm{RL}}  \\ 
&\overset{(a)}{=} \frac{P_\text{s}|h_\mathrm{ST}|^2|h_\mathrm{TL}|^2}{{d}^\chi_\mathrm{ST}{d}^\chi_\mathrm{TL}}
\hspace{-2pt} +\hspace{-2pt}
\frac{P_\mathrm{s}|h_\mathrm{ST}|^2\left|\sum_{n=1}^{N}h_\mathrm{TR_n} h_\mathrm{RL_n}\right|^2}{{d}^\chi_\mathrm{ST}{d}^\chi_\mathrm{TR}{d}^\chi_\mathrm{RL}},
\end{align}
where $(a)$ results from ideal phase shifting in the RIS \cite{ref3941} resulting in $\Psi_n= \phi_n-\delta_n-\zeta_{n}$.
The phase information in tags can be measured by multi-phase probing (MPP)-based technique as described in \cite{BTTN-CSI-1}. Therefore, we can write the incident power on the LT at modulation state $\mathrm{i}$ as \eqref{p_inc_k}.
\begin{align} \label{p_inc_k}
P_\mathrm{inc,i} \hspace{-2pt}=\hspace{-2pt} P_\mathrm{s} \Gamma_\mathrm{L,i}^* \hspace{-2pt} \left( \hspace{-2pt} \frac{|h_\mathrm{ST}|^2|h_\mathrm{TL}|^2}{{d}^\chi_\mathrm{ST}{d}^\chi_\mathrm{TL}} \hspace{-2pt}+\hspace{-2pt} \frac{|h_\mathrm{ST}|^2\left|\sum_{n=1}^{N}h_\mathrm{TR_n} h_\mathrm{RL_n}\right|^2}{{d}^\chi_\mathrm{ST}{d}^\chi_\mathrm{TR}{d}^\chi_\mathrm{RL}}\right).
\end{align}

The energy harvesting circuit uses a two-stage Dickson charge pump for voltage multiplication as shown in Fig. \ref{fig:EH_Circuit} \cite{design-BTTN-3}. 
The first stage of the Dickson charge pump rectifies the input RF signal \( V_\mathrm{rf} \). The voltage across the capacitor \( C_\mathrm{c} \) after the first stage is approximately \( V_\mathrm{rf} - V_\mathrm{d} \), where \( V_\mathrm{d} \) is the forward voltage drop of the diodes.
The second stage further multiplies the voltage. The output voltage after the second stage, considering the rectification, is approximately \( 2 \times (V_\mathrm{rf} - V_\mathrm{d}) \).
Since the circuit has two stages of voltage multiplication, the final output voltage \( V_\text{o} \) is:
\begin{align}
   V_\text{o} \approx 2 \times (2 \times (V_\text{rf} - V_\text{d})) = 4 \times (V_\text{rf} - V_\text{d}) .
\end{align}
By redefining the output voltage \( V_\text{o} \) and the rectified voltage \( V_\mathrm{rf} \) in the energy harvesting circuit for different modulation states as \( V_\mathrm{out,i}^\mathrm{Hrv} \) and \( V_\mathrm{rect,i}^\mathrm{Hrv} \), respectively, we have:
\begin{align} \label{Vo_hrv_01}
   V_\mathrm{out,i}^\mathrm{Hrv} \approx 4 V_\mathrm{rect,i}^\mathrm{Hrv} - 4 V_\text{d}.
\end{align}

The demodulator circuit involves a rectifier followed by a filter and voltage divider network. 
The input RF signal \( V_\text{rf} \) as shown in Fig. \ref{fig:Dem_Circuit} is rectified by the diodes \( D_1 \) and \( D_2 \), resulting in a peak rectified voltage of \( V_\text{rf} - 2V_\text{d} \), considering the forward voltage drops of the diodes.
The rectified voltage is filtered by capacitors \( C_3 \) and \( C_4 \) to smooth out the ripples, providing a DC voltage \( V_\text{baseband} \) approximately equal to \( V_\text{rf} - 2V_\text{d} \).
The filtered DC voltage is then divided by the resistors \( R_1 \) and \( R_2 \). The voltage across \( R_2 \), which we denote as \( V_\text{o} \), is given by the voltage divider formula:
\begin{align}
    V_\text{o} = V_\text{baseband} \cdot \frac{R_2}{R_1 + R_2} ,
\end{align}
where \( V_\text{baseband} \approx V_\text{rf} - 2V_\text{d} \). Therefore, we have
\begin{align}
    V_\text{o} \approx (V_\text{rf} - 2V_\text{d}) \cdot \frac{R_2}{R_1 + R_2} .
\end{align}
By redefining the output voltage \( V_\text{o} \) and the rectified voltage \( V_\mathrm{rf} \) in the demodulator circuit for different modulation states as \( V_\mathrm{out,i}^\mathrm{Dem} \) and \( V_\mathrm{rect,i}^\mathrm{Dem} \), respectively, we have
\begin{align} \label{Vo_dem_01}
    V_\mathrm{out,i}^\mathrm{Dem} \approx \left(V_\mathrm{rect,i}^\mathrm{Dem} - 2V_\text{d}\right) \cdot \frac{R_2}{R_1 + R_2}.
\end{align}

Let's say the incident power at the tag is divided between the energy harvesting and demodulator circuits based on the coefficient \( \alpha \) as \eqref{hrv_alfa} and\eqref{dem_alfa}, respectviely. 
\begin{align} \label{hrv_alfa}
    P_\text{inc,i}^\text{Hrv} = \alpha \  P_\text{inc,i},
\end{align} 
\begin{align} \label{dem_alfa}
    P_\text{inc,i}^\text{Dem} = (1 - \alpha) \ P_\text{inc,i},
\end{align}
As the squared voltage at the output of the energy harvesting and AM demodulator (typically an envelope detector) within the tag’s integrated circuit (IC) is directly proportional to the power absorbed by the chip \cite{passive-BTTN-1}, the rectified voltage for the energy harvesting and demodulator circuits in the $\text{on}$ and $\text{off}$ states can be written as
\begin{align} \label{V_rect_hrv}
    V_\text{rect,i}^\text{Hrv} = \sqrt{k_\text{Hrv} P_\text{inc,i}^\text{Hrv}},
\end{align}
\begin{align} \label{V_rect_dem}
    V_\text{rect,i}^\text{Dem} = \sqrt{k_\text{Dem} P_\text{inc,i}^\text{Dem}}.
\end{align}
By inserting \eqref{V_rect_hrv} and \eqref{hrv_alfa} into \eqref{Vo_hrv_01}, we rewrite the output voltage in the energy harvesting circuit as: 
\begin{align} \label{Vo_hrv_02}
    V_\text{out,i}^\text{Hrv} \approx 4 \sqrt{k_\text{Hrv} \ \alpha \ P_\text{inc,i}} - 4 V_\text{d}.
\end{align}
Now, by substituting \eqref{p_inc_k} into \eqref{Vo_hrv_02}, we have the output
voltage in the energy harvesting circuit as \eqref{Vo-hrv}. 
\begin{figure*}[t]
			\normalsize
\begin{align}  \label{Vo-hrv}
V_\mathrm{out,i}^\mathrm{Hrv} \approx 4 \sqrt{k_\mathrm{Hrv} \ \alpha \ P_\mathrm{s} \ \Gamma_\mathrm{L,i}^* \left( \frac{|h_\mathrm{ST}|^2|h_\mathrm{TL}|^2}{{d}^\chi_\mathrm{ST}{d}^\chi_\mathrm{TL}} + \frac{|h_\mathrm{ST}|^2\left|\sum_{n=1}^{N}h_\mathrm{TR_n} h_\mathrm{RL_n}\right|^2}{{d}^\chi_\mathrm{ST}{d}^\chi_\mathrm{TR}{d}^\chi_\mathrm{RL}}\right)} - 4 V_\text{d}.  
\end{align}
\hrulefill
\vspace{-8pt}
\end{figure*}

In a same way, by inserting \eqref{V_rect_dem} and \eqref{dem_alfa} into \eqref{Vo_dem_01}, we rewrite the output voltage in the demodulator circuit as: 
\begin{align} \label{Vo_dem_02}
    V_\text{out,i}^\text{Dem} \approx \left( \sqrt{k_\text{Dem} \ (1-\alpha) \ P_\text{inc,i}} - 2 V_\text{d} \right) \frac{R_2}{R_1 + R_2}.
\end{align}
Now, by substituting \eqref{p_inc_k} into \eqref{Vo_dem_02}, we have the output voltage in the demodulator circuit as \eqref{Vo-dem}. 
\begin{figure*}[t]
			\normalsize
\begin{align}  \label{Vo-dem}
 V_\mathrm{out,i}^\mathrm{Dem} \approx \left( \sqrt{k_\mathrm{Dem} (1 - \alpha) P_\mathrm{s} \ \Gamma_\mathrm{L,i}^* \left( \frac{|h_\mathrm{ST}|^2|h_\mathrm{TL}|^2}{{d}^\chi_\mathrm{ST}{d}^\chi_\mathrm{TL}} + \frac{|h_\mathrm{ST}|^2\left|\sum_{n=1}^{N}h_\mathrm{TR_n} h_\mathrm{RL_n}\right|^2}{{d}^\chi_\mathrm{ST}{d}^\chi_\mathrm{TR}{d}^\chi_\mathrm{RL}}\right) } - 2V_\text{d} \right) \frac{R_2}{R_1 + R_2} . 
\end{align}
\hrulefill
\end{figure*}

\subsection{Initialization Phase}

In practical BTTN deployments, each tag must be registered within the system using a verified and unique identity, making the initialization phase critical to ensure secure communications. 
This stage mainly involves two key steps: the registration of the tags and to the system and the creation of a baseline voltage profile for future authentication. This phase sets up the fundamental parameters and configurations required for the authentication process.

\subsubsection{Tag Registration} 
In the registration step of the studied system model in this paper, TT and LT register with a trusted central server. The server acts as the authority that manages the identification and initial configuration of the tags. 
Depending on the application context, this server could be a smart home IoT hub, a smart factory edge server, or a dedicated IoT gateway in other environments \cite{Multiprotocol-Backscatter}.
TT and LT transmit their unique identifiers, such as pre-programmed serial numbers, to the server through a secure communication channel by modulating their load impedance and backscattering the RF signal.
The server may also configure the RIS by accessing to its microcontroller with the relevant settings and ensure that it recognizes TT and LT as legitimate entities in the system. The registration process ensures that both tags are properly authenticated and ready for secure interactions in the RIS-aided BTTN environment.

\subsubsection{Baseline Voltage Profile Creation}
Following registration, LT establishes a baseline voltage profile to enable future TT authentication. This step begins with TT transmitting a series of known pilot signals while modulating the RF field through distinct modulation states, such as ‘on’ and ‘off’.
To enhance the quality of these signals, RIS is configured adaptively to optimize the RF energy distribution between TT and LT. 
The RIS microcontroller determines the phase-shifting coefficients adaptively based on the received pilot signals, ensuring maximum signal strength at LT. This configuration requires the availability of CSI for the cascade links, which is either locally estimated at the RIS microcontroller or provided by the server, depending on the system design. If provided by the server, the RIS microcontroller adjusts the reflective elements dynamically according to the CSI updates received. Periodic recalibration may also be performed if environmental conditions change.

LT then measures the output voltage generated by its energy harvesting and demodulation circuits for each modulation state, capturing the unique signal characteristics enhanced by the RIS configuration to establish a robust baseline voltage profile.
The measured voltages, denoted as \( V_\mathrm{out,0}^\mathrm{Hrv} \) and \( V_\mathrm{out,0}^\mathrm{Dem} \), correspond to TT's unique physical layer signature and can be aproximated as shown in \eqref{Vo-hrv} and \eqref{Vo-dem}, respectively. These profiles together with TT's identifier ($ID_\mathrm{T}$) are securely stored in LT's memory for comparison during future authentication sessions. 
Optionally, the central server may also maintain a backup of these profiles to ensure system resilience against memory loss or corruption in LT.
\begin{figure}[t]
    \centering    \includegraphics[width=0.48\textwidth]{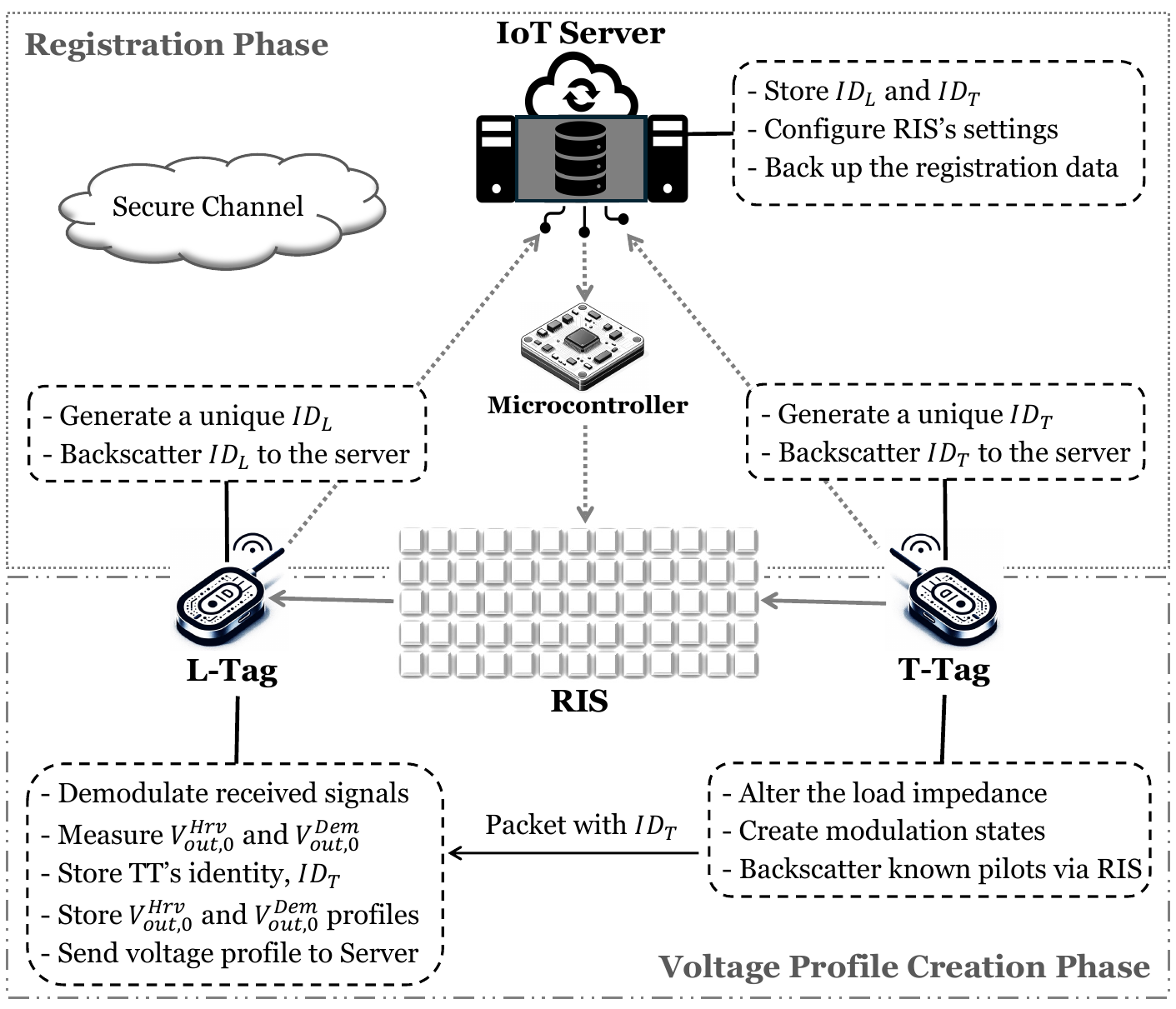}
    \caption{Different steps of the initialization phase.}
    \label{fig-init}
\end{figure}
It should be noted that to prevent adversaries from interfering with the initialization phase, all communication between TT, LT, RIS, and the server must be conducted over a secure channel. Additionally, the environment in which the baseline voltage profiles are measured must be free from interference or environmental noises to ensure the integrity of the recorded profiles. The trusted server ensures that all entities, including RIS, operate securely and under legitimate control during this phase.
By completion of the initialization phase, TT and LT are securely registered, and LT is equipped with a reliable baseline voltage profile to authenticate TT in subsequent interactions.
Fig. \ref{fig-init} illustrates the various stages of the initialization phase.

\subsection{Authentication Phase}

Once the initialization phase is complete, TT and LT enter the operational stage, where TT’s identity is verified by LT during each communication session. This phase involves the following key steps:

\subsubsection{Backscattering Signal by TT}
TT initiates the authentication process by backscattering its modulated signal, which encodes its unique identifier \( ID_T \) along with session-specific data. To enhance security, TT modulates the RF field in a manner consistent with its registered baseline voltage profile, cycling through predefined modulation states (e.g., on and off) similar to the pilot signals transmitted during initialization. This ensures that the transmitted signal maintains the physical layer characteristics expected by LT.

\subsubsection{Signal Reception and Voltage Measurement by LT}
LT receives the backscattered signal from TT and extracts the transmitted identifier \( ID_T \). Simultaneously, LT measures the output voltage generated by its energy harvesting and demodulation circuits for each modulation state. These measured voltages, \( V_\mathrm{out}^\mathrm{Hrv} \) and \( V_\mathrm{out}^\mathrm{Dem} \), represent TT’s current physical layer signature under operational conditions.
The uniqueness of these voltage profiles is intrinsically tied to TT’s location-specific CSI, ensuring that the physical layer characteristics of the backscattered signal are distinct for a tag at a particular location. Additionally, this uniqueness is further influenced by the physical circuitry of LT, including the capacitance and resistance of the capacitors, resistors, and diodes in its energy harvesting and demodulation circuits.
During this step, the RIS dynamically adjusts its phase-shifting coefficients ($\Psi$) to optimize the received signal strength at LT. By fine-tuning the phase shifts of its reflective elements, the RIS maximizes the SNR at LT, which improves the reliability and precision of the voltage measurements. 

\subsubsection{Profile Comparison and Decision}
LT compares the measured voltages \( V_\mathrm{out}^\mathrm{Hrv} \) and \( V_\mathrm{out}^\mathrm{Dem} \) with the baseline profiles \( V_\mathrm{out,0}^\mathrm{Hrv} \) and \( V_\mathrm{out,0}^\mathrm{Dem} \) stored during initialization. The comparison is conducted as follows:
\begin{align}
    \Delta V_\mathrm{Hrv} &= \left|V_\mathrm{out}^\mathrm{Hrv} - V_\mathrm{out,0}^\mathrm{Hrv}\right|, \\
    \Delta V_\mathrm{Dem} &= \left|V_\mathrm{out}^\mathrm{Dem} - V_\mathrm{out,0}^\mathrm{Dem}\right|.
\end{align}
If \( \Delta V_\mathrm{Hrv} \) and \( \Delta V_\mathrm{Dem} \) fall within pre-defined thresholds \( \tau_\mathrm{Hrv} \) and \( \tau_\mathrm{Dem} \), respectively, LT authenticates TT as a legitimate tag. Otherwise, the authentication is rejected.
Within the tag's circuitry, this comparison process is facilitated by a simple comparator unit. The comparator takes the measured voltages and baseline profiles as inputs and calculates the voltage differences \( \Delta V_\mathrm{Hrv} \) and \( \Delta V_\mathrm{Dem} \). 
To determine the legitimacy of TT during the authentication phase, a binary hypothesis test can be conducted based on the deviation of the measured voltages from their corresponding baseline values as follows:
\begin{equation}
\begin{aligned} \label{hypo-eq}
\begin{cases} 
\mathcal{H}_0: & \Delta V_\mathrm{Hrv} \leq \tau_\mathrm{Hrv} \ \ \mathrm{and} \ \ \Delta V_\mathrm{Dem} \leq \tau_\mathrm{Dem}, \\
\mathcal{H}_1: & \Delta V_\mathrm{Hrv} > \tau_\mathrm{Hrv} \ \ \ \mathrm{or} \ \ \ \Delta V_\mathrm{Dem} > \tau_\mathrm{Dem},
\end{cases}
\end{aligned}
\end{equation}
where $\mathcal{H}_0$ indicates the case that the measured voltage profiles \( V_\mathrm{out}^\mathrm{Hrv} \) and \( V_\mathrm{out}^\mathrm{Dem} \) belong to a legitimate TT while $\mathcal{H}_0$ indicates that an adversarial attempt might happen. 
This implementation ensures low power consumption and quick decision-making, aligning with the resource constraints of passive BTTN tags. 

\subsubsection{Authentication Confirmation} 
\begin{figure}[t]
    \centering
    \includegraphics[width=0.49\textwidth]{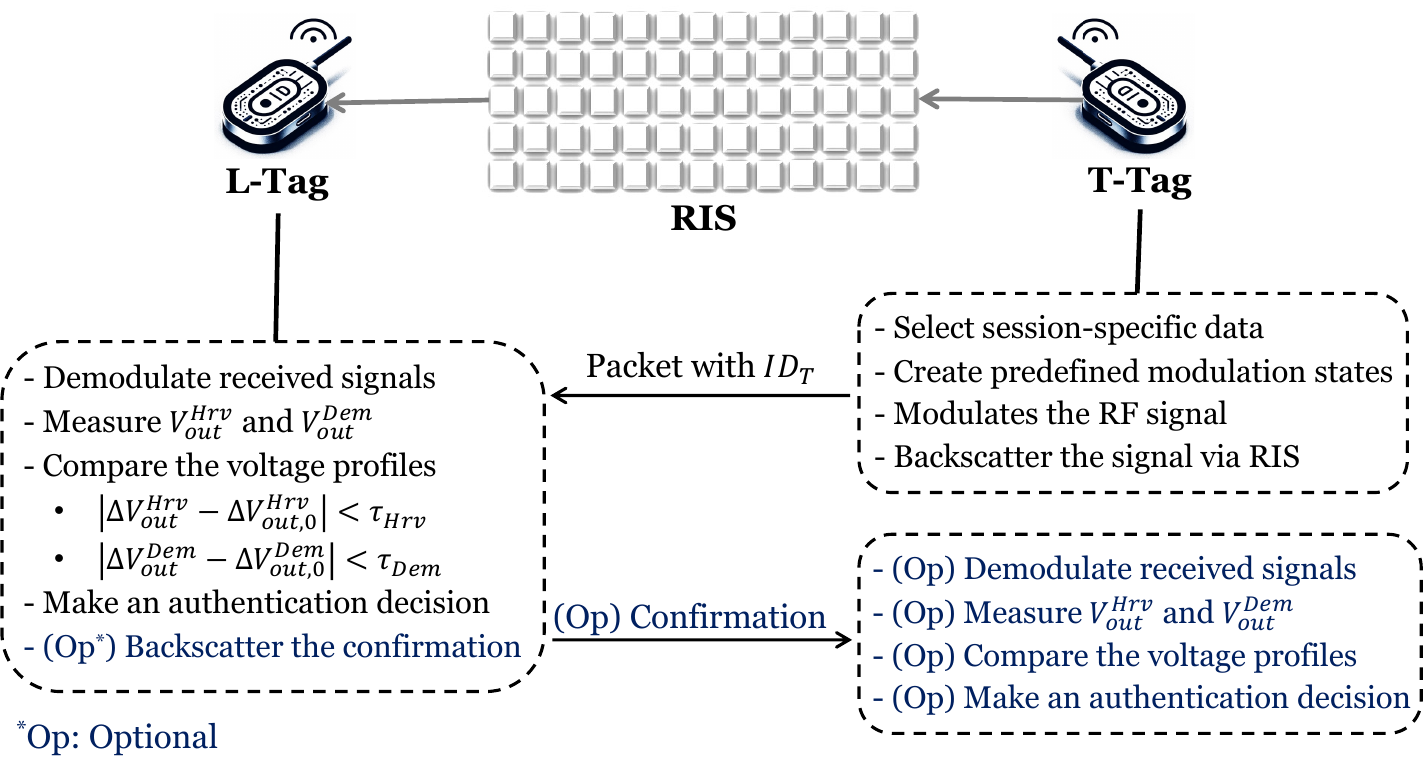} 
    \caption{Different steps of the authentication phase.}
    \label{fig:auth_phase}
\end{figure}

Upon successful authentication, LT can optionally transmit a confirmation message to TT, allowing secure data exchange to proceed. This confirmation message can include additional session-specific parameters to establish a secure communication session.
Following this, TT can authenticate LT in an exact same manner by leveraging the baseline voltage profiles \( V_\mathrm{out,0}^\mathrm{Hrv} \) and \( V_\mathrm{out,0}^\mathrm{Dem} \) established during the initialization phase. TT measures the output voltages generated by its circuits upon receiving backscattered signals from LT and compares them against the baseline profiles stored in its memory. By repeating the same process of voltage profile matching and threshold validation, TT ensures that the backscattered signal originates from LT.
This mutual authentication process ensures bidirectional security, safeguarding both tags against impersonation attacks, and establishing a trusted environment for subsequent data exchange. 
Any authentication failure at any of these stages will let the tags discard the received signal and may alert the central server for anomaly detection and system diagnostics.
Fig. \ref{fig:auth_phase} shows different steps in the authentication phase.
\begin{remark}
The effectiveness of the proposed authentication phase is fundamentally dependent on the integrity of the baseline voltage profiles, \( V_\mathrm{out,0}^\mathrm{Hrv} \) and \( V_\mathrm{out,0}^\mathrm{Dem} \), established during the initialization phase. Reliable physical measurements from the energy harvesting and demodulation circuits are crucial for ensuring accurate and consistent voltage profiles. However, various factors can affect the reliability of these measurements. Environmental conditions, such as temperature fluctuations, humidity, or RF signal obstructions, may introduce noise or distortions, leading to deviations in the recorded profiles. Similarly, hardware limitations or variations in circuit components, including diodes, capacitors, or resistors, can cause nonlinearities that degrade the quality of the measured voltages. 
Moreover, the uniqueness of the physical layer attributes, which forms the foundation of the authentication scheme, is inherently tied to the CSI of the direct and cascade links. These coefficients encapsulate the unique propagation characteristics of the RF signal between the TT and LT, influenced by factors such as distance, multipath fading and shadowing effects, and obstacles. This dependency ensures that the physical-layer features are specific to the tag's location and cannot be easily replicated by an adversary from a different location.
\end{remark}
\begin{remark}
The RIS configuration plays a critical role in ensuring high-quality signal strength at the LT. A well-optimized and secure RIS controller is essential for enhancing the SNR, thereby improving the accuracy of the voltage profiles. 
Additionally, the proper selection of threshold values \( \tau_\mathrm{Hrv} \) and \( \tau_\mathrm{Dem} \) is pivotal in determining the authentication system's performance. Thresholds must be carefully tuned to balance the trade-off between false acceptance rates (FAR) and false rejection rates (FRR). Overly strict thresholds could result in legitimate devices being rejected, while overly lenient thresholds might leave the system vulnerable to adversarial attacks. 
\end{remark}

\section{Security Analysis} \label{sec_analy}

This section presents a security analysis of the proposed authentication scheme, focusing on its resilience to potential threats encountered in RIS-aided BTTNs. 

\subsection{Impersonation Attacks}

In this attack scenario, Eve attempts to masquerade as a legitimate TT by replicating its identifier and physical layer attributes to pass the authentication query at LT. 
The proposed PLA's resistance to impersonation attack is rooted in the uniqueness of the CSI between TT and LT, which underpins the generation of location-specific voltage profiles in LT, \( V_\mathrm{out}^\mathrm{Hrv} \) and \( V_\mathrm{out}^\mathrm{Dem} \), as approximated in \eqref{Vo-hrv} and \eqref{Vo-dem}, respectively. 
During impersonation, Eve generates a backscattered signal with \( ID_T \), resulting in:
\begin{align} \label{Y-Eve}
    Y_\mathrm{L}^\mathrm{E} \hspace{-2pt}=\hspace{-2pt} \mathcal{B}_\mathrm{E} S(t) \sqrt{P_\mathrm{s}} \bigg(\hspace{-2pt} h_\mathrm{SE} h_\mathrm{EL} \hspace{-2pt}+\hspace{-2pt} h_\mathrm{SE} \mathbf{H}_\mathrm{RL}^\mathcal{T} \mathbf{\Phi}_\mathrm{E} \mathbf{H}_\mathrm{ER} \hspace{-2pt}\bigg) \hspace{-2pt}+\hspace{-2pt} n_\mathrm{L},
\end{align}
where \( \mathbf{\Phi}_\mathrm{E} =\text{diag}\left(\left[\mathrm{e}^{j\phi_\mathrm{e_1}}, \mathrm{e}^{j\phi_\mathrm{e_2}},...,\mathrm{e}^{j\phi_\mathrm{e_N}}\right]\right) \) refers to a suboptimal RIS configuration, as Eve lacks access to the RIS microcontroller and her CSI is also unknown by RIS as she is a passive attacker. \( h_\mathrm{SE} \) and \( h_\mathrm{EL} \) shows the source-to-Eve and Eve-to-LT channel coefficients, respectively, and \(\mathbf{H}_\mathrm{ER}\) denotes channel coefficients between Eve and RIS. 
Comparing to \eqref{eq-LT-sig}, the spatial separation between TT and Eve, which is typically greater than half a wavelength, ensures that the CSI at LT for TT cannot be replicated by Eve. 
Moreover, the received power at LT, \( P_\mathrm{inc} \), is predominantly influenced by the cascade link facilitated by RIS, which directly impacts the voltage profiles measured by LT’s energy harvesting and demodulation circuits. 
Since Eve lacks access to the RIS microcontroller and cannot manipulate its configuration, she is unable to replicate the RIS-enhanced signal propagation necessary to recreate TT’s voltage profiles at LT. 
As a result, the voltage profiles generated by LT in response to Eve’s backscattered signals shown in \eqref{Y-Eve} will differ significantly from the baseline profiles stored during initialization.
The authentication process further incorporates binary hypothesis testing, as outlined in \eqref{hypo-eq}, to evaluate the differences between the measured voltages during authentication and the baseline profiles. By setting appropriate thresholds, \( \tau_\mathrm{Hrv} \) and \( \tau_\mathrm{Dem} \), LT can effectively reject any authentication attempts where these differences exceed the defined thresholds, ensuring that impersonation attempts are thwarted. 

\subsection{MITM Attack}

In an MITM attack on BTTNs, Eve strategically positions herself between TT and LT, intercepting and potentially manipulating the signals exchanged between legitimate endpoints. The goal of such an attack is to eavesdrop on the communication or inject maliciously altered signals while deceiving LT into believing the signals originate from TT.
In the proposed PLA, the voltages measured at LT's energy harvesting and demodulator circuits 
are derived from the real-time CSI between TT and LT, as described in \eqref{Vo-hrv} and \eqref{Vo-dem}. When Eve manipulates the intercepted signal to LT, the CSI between Eve and LT and consequently the received power $P_\mathrm{inc}$ differs significantly from of the original TT-to-LT channel. 
This disparity becomes even more pronounced with the integration of RIS into the system. As RIS is exclusively configured to optimize the TT-LT link, the voltages measured at LT during the attack, \( V_\mathrm{out,MITM}^\mathrm{Hrv} \) and \( V_\mathrm{out,MITM}^\mathrm{Dem} \), show a greater deviation from the expected baseline profiles \( V_\mathrm{out,0}^\mathrm{Hrv} \) and \( V_\mathrm{out,0}^\mathrm{Dem} \).
Using predefined voltage thresholds \( \tau_\mathrm{Hrv} \) and \( \tau_\mathrm{Dem} \) to evaluate these deviations through the hypothesis testing framework described in \eqref{hypo-eq} during the authentication process, 
LT ensures that any manipulated or intercepted signals are detected and discarded.

\subsection{Replay Attacks}

Replay attacks occur when Eve records legitimate authentication signals transmitted by TT during a valid session and replays them to deceive LT into granting authentication. Such attacks exploit the static nature of backscattered signals, where no time-varying cryptographic keys or dynamic identifiers are employed in passive and ultra-low power BTTN communication systems. The proposed PLA is inherently resistant to replay attacks due to the reliance on dynamic, location-specific physical layer characteristics and real-time voltage profile measurements at LT.
In the proposed system, LT authenticates TT based on the unique voltage profiles, \( V_\mathrm{out}^\mathrm{Hrv} \) and \( V_\mathrm{out}^\mathrm{Dem} \), which are directly influenced by the real-time CSI between TT and LT, as well as the RIS-enhanced signal propagation. These voltage profiles are location-specific, reflecting the current state of the RIS configuration and the RF environment. When Eve replays previously recorded signals, the CSI between Eve and LT will differ from the original TT-to-LT link due to spatial and temporal variations in the wireless channel. Consequently, the voltages measured at LT during the replay attack will deviate from the baseline profiles established during initialization.
This ensures that the replayed signals, even if identical to the original, fail to produce the same voltage profiles at LT, providing robust protection against replay attacks.

\subsection{Relay Attacks}

Relay attacks involve Eve intercepting communication signals between TT and LT and relaying them to simulate legitimate interactions. In the studied RIS-aided BTTN system, Eve can attempt to relay the signal from TT to LT by capturing \( Y_\mathrm{L} \) and retransmitting it. 
%
%
The channel coefficients between Eve and LT, denoted as \( h_\mathrm{EL} \) and \(\mathbf{H}_\mathrm{ER}\), differs from the ones between TT and LT. 
Moreover, the RIS-enhanced signal propagation optimized for TT cannot benefit Eve, as the RIS configuration (\( \mathbf{\Phi} \)) is determined by the CSI of the legitimate TT-LT link.
Consequently, the physical voltage profiles measured at LT during the relay attack, \( V_\mathrm{out,Rel}^\mathrm{Hrv} \) and \( V_\mathrm{out,Rel}^\mathrm{Dem} \), will deviate from the baseline profiles \( V_\mathrm{out,0}^\mathrm{Hrv} \) and \( V_\mathrm{out,0}^\mathrm{Dem} \), which were established during the initialization phase. 
The authentication process evaluates the differences between the measured and baseline voltage profiles using the hypothesis test outlined in \eqref{hypo-eq}. Specifically:
\begin{align}
    \Delta V_\mathrm{Hrv}^\mathrm{Rel} &= \left|V_\mathrm{out,Rel}^\mathrm{Hrv} - V_\mathrm{out,0}^\mathrm{Hrv}\right|, \\
    \Delta V_\mathrm{Dem}^\mathrm{Rel} &= \left|V_\mathrm{out,Rel}^\mathrm{Dem} - V_\mathrm{out,0}^\mathrm{Dem}\right|.
\end{align}
%
Since Eve cannot produce the location-specific CSI of TT and the RIS is not optimized for her signal, it is highly unlikely that \( \Delta V_\mathrm{Hrv}^\mathrm{Rel} \) or \( \Delta V_\mathrm{Dem}^\mathrm{Rel} \) would fall within the thresholds \( \tau_\mathrm{Hrv} \) or \( \tau_\mathrm{Dem} \). Therefore, the system can effectively detect and thwart relay attacks, ensuring the robustness of the authentication framework.

\section{Performance Evaluation} \label{sec_results} 

In this section, we assess the performance of the proposed PLA through simulations conducted under various system configurations and threat scenarios. The evaluation considers both RIS-enabled and non-RIS BTTNs, focusing on the accuracy and robustness of the authentication scheme. Specifically, we analyzed the impact of varying tag-to-tag and Eve-to-tag distances, different RF power levels, and varying numbers of RIS elements under multiple threat conditions.

\subsection{Experimental Setting}

\subsubsection{Simulation Setup}
For our simulation setup, we considered an indoor RIS-aided BTTN environment, such as a smart home, with fixed positions for the RF source, tags, RIS, and Eve, to replicate realistic communication scenarios, where LT is exposed to various attack vectors. 
The RIS is strategically placed to enhance the received signal power at the tags, while Eve is positioned between TT and LT to execute her malicious activities. This arrangement facilitates a detailed evaluation of PLA performance with predefined distances: \(d_\mathrm{ST} = d_\mathrm{SL} = 1\)m, \(d_\mathrm{TL} = 1.5\)m and \(d_\mathrm{TR} = d_\mathrm{RL} = 1\)m, and Eve being placed slightly closer to the RIS for a worst-case scenario at \(d_\mathrm{LE} = d_\mathrm{TE} = 75\) cm and \(d_\mathrm{ER} = 70\) cm, making it well suited for indoor RIS-aided BTTN applications. Additional simulation parameters include a spectral efficiency of \(R_\mathrm{s} = 1\) bps/Hz, noise power levels of \(\sigma^2_\mathrm{T} = \sigma^2_\mathrm{L} = -40\) dBm at the tags and \(\sigma^2_\mathrm{E} = -30\) dBm at Eve, a maximum average source power of \(P_s = 1\) dBm, and path loss exponents of \(\chi_1 = 3.5\) for direct links and \(\chi_2 = 2.5\) for RIS-aided links.
The simulations are conducted using MATLAB 2023a on a MacBook Air equipped with an Apple M1 chip and 16 GB of RAM. Two distinct scenarios are analyzed: a baseline BTTN configuration without RIS and a RIS-enhanced BTTN setup with \(N = 20, 50, 100\) reflective elements. 
The communication channels between all entities are modeled using the Rician fading model, which is suitable for the short-range application scenarios in this paper \cite{BC_Rician}. The channel variance for subchannel \(C_\mathrm{k}\) is expressed as:
\begin{align}
    \sigma_\mathrm{h_\mathrm{TL}} = \left(\frac{\lambda}{4\pi d_\mathrm{TL}} G_\mathrm{T} G_\mathrm{L}\right),
\end{align}
where \(\lambda = \frac{c}{C_\mathrm{X}}\) represents the signal wavelength, \(c\) is the speed of light, and \(G_\mathrm{X} = 8\), denoting the antenna gain of $\mathrm{X}$ with \(\mathrm{X} \in \{\mathrm{T}, \mathrm{L}\}\) \cite{PLA-MCBF}.
\subsubsection{Performance Metrics}
To evaluate the performance of the proposed authentication scheme, we employ several metrics with an emphasis on the \textit{accuracy} and \textit{robustness} of the authentication, where the first offers valuable insight into the system's ability to accurately identify legitimate endpoints and the latter represents the capacity of system to resist against various attacks by effectively rejecting unauthorized access attempts. 
During the authentication process (where LT authenticates TT), the outcomes are categorized as $Accept$ or $Reject$ according to the hypothesis test shown in \eqref{hypo-eq}. To measure the performance of our PLA method, we use the following metrics: 

\textit{a) Authentication Rate}, which reflects accurate acceptance of legitimate tags or, equivalently, the true positive rate (TPR), which is defined as:
\begin{align}
 \small   TPR = \frac{N_\mathrm{TP}}{N_\mathrm{TP} + N_\mathrm{FN}},
\end{align}
where \( N_\mathrm{TP} \) and \( N_\mathrm{FN} \) represent the number of times TT is correctly accepted and incorrectly rejected, respectively.

\textit{b) False Acceptance Rate}, signifying the false positive rate (FPR), which quantifies the rate at which Eve is mistakenly accepted by the system and can be defined as:
\begin{align}
   \small  FPR = \frac{N_\mathrm{FP}}{N_\mathrm{FP} + N_\mathrm{TN}},
\end{align}
where \( N_\mathrm{FP} \) and \( N_\mathrm{TN} \) represent the number of times Eve was incorrectly accepted and correctly rejected, respectively.
Furthermore, we used the Receiver Operating Characteristic (ROC) curve to examine the relationship between the authentication rate and false acceptance rate under various thresholds, threat scenarios, and system configurations.

\subsection{Simulation Results and Discussion}
We further analyze the performance of the proposed PLA, focusing on its accuracy and robustness across different system configurations and attack scenarios.
\subsubsection{BTTN Without RIS}
\begin{table}[] 
\centering
\caption{Authentication Rate vs. Different TT-to-LT Distance for Selected FPR Values}
\begin{tabular}{lc|c|c|c|c|}
\cline{3-6}
                                           & \multicolumn{1}{l|}{} & $d_\mathrm{TL}\hspace{-2pt}=\hspace{-2pt}0.5$m & $d_\mathrm{TL}\hspace{-2pt}=\hspace{-2pt}1$m & $d_\mathrm{TL}\hspace{-2pt}=\hspace{-2pt}1.5$m & $d_\mathrm{TL}\hspace{-2pt}=\hspace{-2pt}2$m \\ \hline
\multicolumn{1}{|l|}{\multirow{4}{*}{\rotatebox[origin=c]{90}{FPR}}} & 0.15  & 96.29\%    & 91.88\%  & 86.01\%    & 71.89\%  \\ \cline{2-6} 
\multicolumn{1}{|l|}{}    & 0.20  & 99.52\%    & 96.19\%  & 86.71\%    & 80.23\%  \\ \cline{2-6} 
\multicolumn{1}{|l|}{} & 0.25  & 99.98\%    & 98.79\%  & 94.04\%    & 88.54\%  \\ \cline{2-6} 
\multicolumn{1}{|l|}{}  & 0.30   & 99.99\%    & 99.59\%  & 97.19\%    & 92.61\%  \\ \hline
\end{tabular} \label{Table:d_TL}
\end{table}
Table \ref{Table:d_TL} demonstrates the variation in the authentication rate of the proposed PLA with $d_\mathrm{TL}$ across different FPR thresholds.
A clear trend emerges: as \(d_\mathrm{TL}\) increases, the authentication rate consistently decreases across all FPR values. 
For example, at \(FPR = 0.15\), the authentication rate drops from \(96.29\%\) at \(d_\mathrm{TL} = 0.5 \, \mathrm{m}\) to \(71.89\%\) at \(d_\mathrm{TL} = 2 \, \mathrm{m}\). 
The inherent weakness of backscattered signals in BTTNs, which rely on passive communication and minimal power levels, makes them particularly vulnerable to distance-induced signal degradation and causes the measured voltages to deviate further from the baseline profiles, thereby elevating the likelihood of authentication failures. As a result, maintaining reliable authentication becomes increasingly challenging over extended distances in BTTNs. 
Furthermore, the authentication rate improves when the FPR threshold increases, as a higher FPR threshold allows for a greater margin of deviation between the measured and baseline voltage profiles during authentication. While this increases the chances of accepting legitimate tags, it also comes at the cost of potentially admitting more false positives, although this trade-off is not captured in this table.
\begin{figure}[t]
    \centering
    \includegraphics[width=0.36\textwidth]{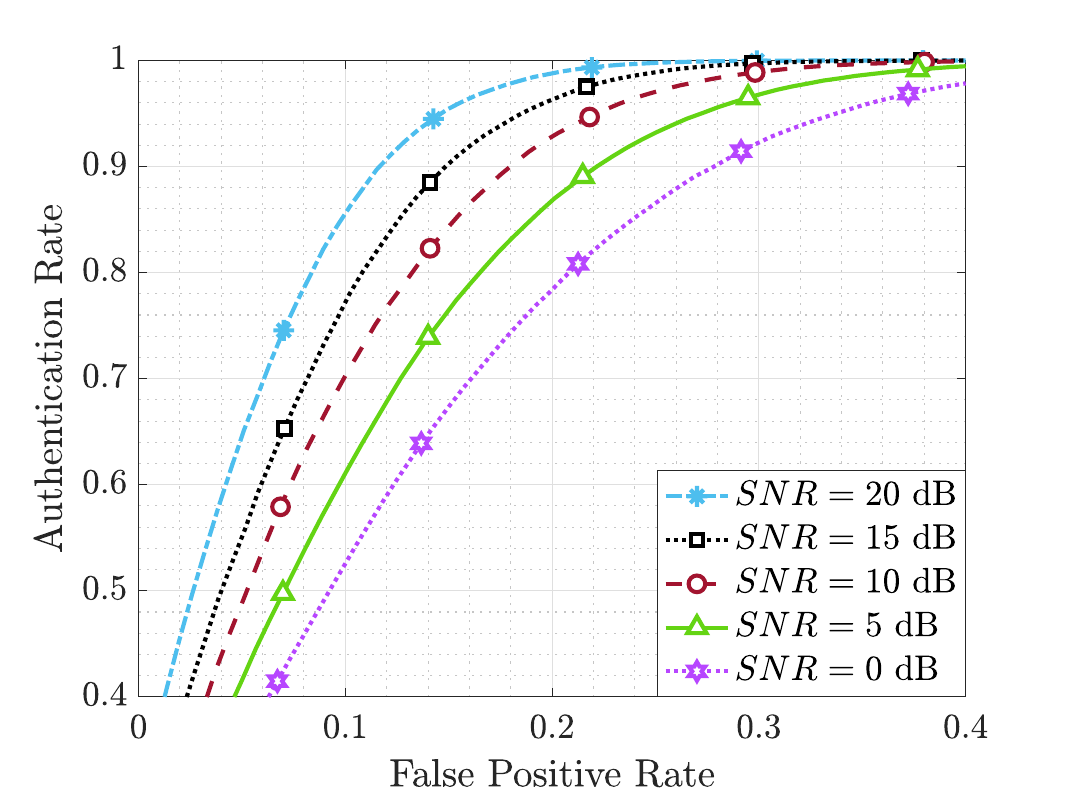} 
    \caption{ROC for TT authenticating LT for different SNR at LT.}
    \label{fig:Tag_SNR}
\end{figure}
Fig. \ref{fig:Tag_SNR} illustrates the authentication rate under varying SNR levels at LT. It can be observed that higher SNR values enhance the authentication rate for any given FPR. 
This improvement is due to the stronger reception of the signal in LT, which enhances the reliability of the measured voltage profiles, 
thus reducing the probability of errors during the authentication process. 
The results in Fig. \ref{fig:Tag_SNR} highlight the critical role of signal quality in ensuring robust and reliable authentication in BTTNs, where passive tags are inherently limited in their power and processing capabilities.

\subsubsection{RIS-aided BTTN}
\begin{figure}[t]
    \centering
    \includegraphics[width=0.36\textwidth]{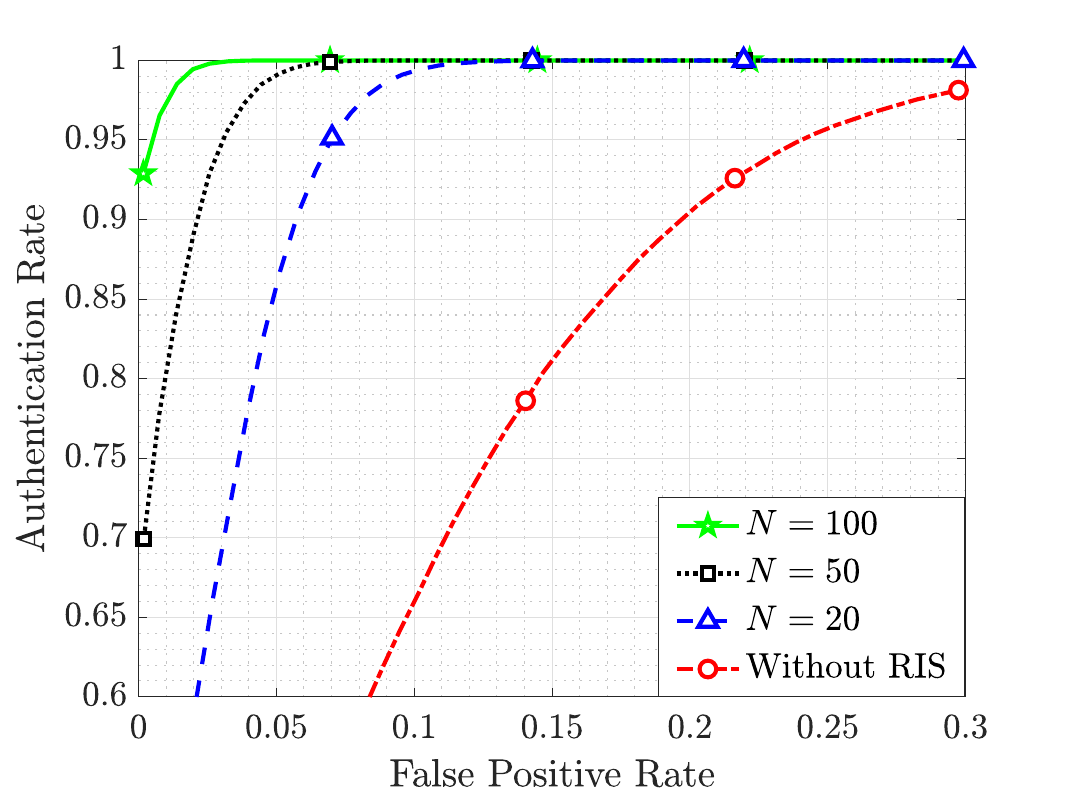} 
    \caption{ROC for TT authenticating LT for different RIS scenarios and number of reflective elements.}
    \label{fig:TPR_N_RIS}
\end{figure}
Fig. \ref{fig:TPR_N_RIS} presents the ROC curves for TT to authenticate LT under a different number of RIS elements and the absence of RIS. The results reveal a substantial enhancement in authentication performance with an increasing number of RIS elements, especially compared to the \textit{Without RIS} scenario. This behavior is attributed to the RIS's ability to optimize the signal propagation between TT and LT by dynamically configuring its phase-shifting elements, thereby improving the received signal strength at LT and enhancing the reliability of the measured voltage profiles, which reduces the deviation from the baseline profiles and increases the probability of accurate authentication. 

\begin{figure}[t]
    \centering
    \includegraphics[width=0.36\textwidth]{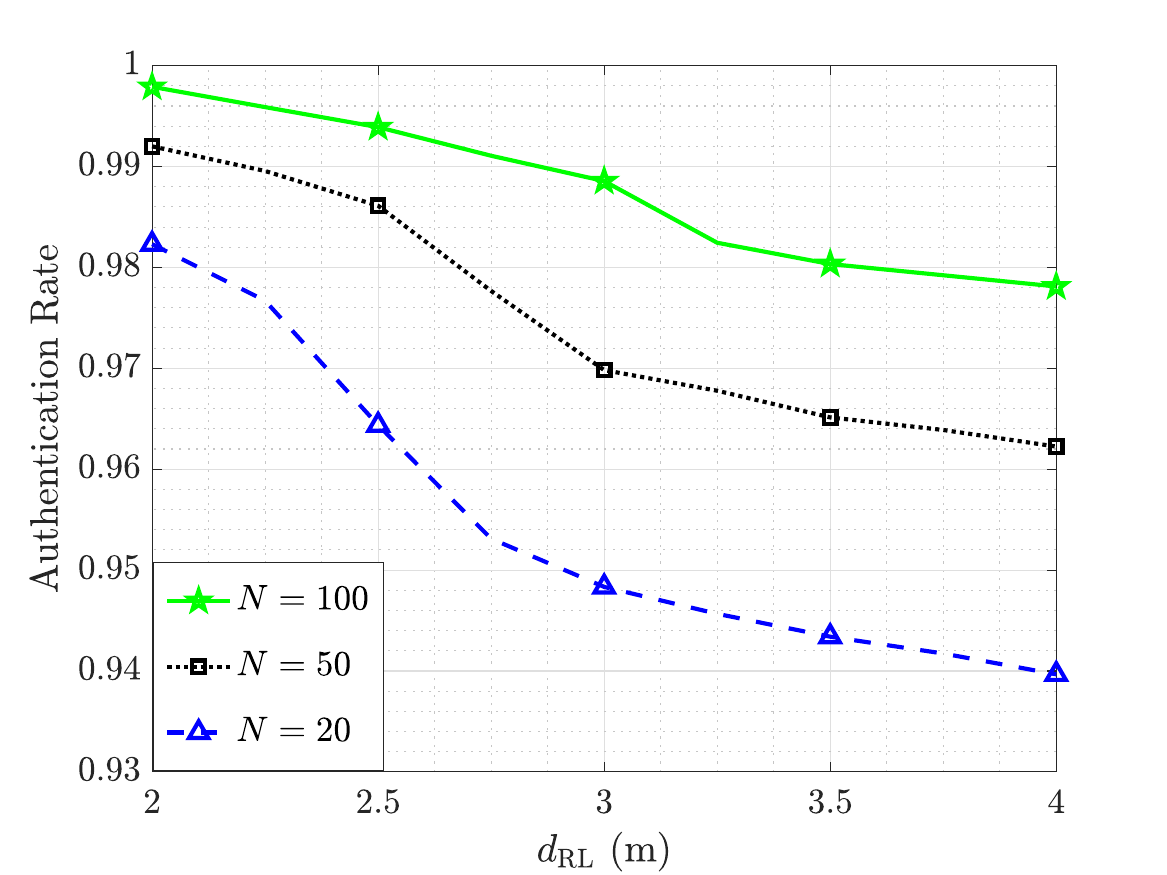} 
    \caption{Impact of $d_\mathrm{RL}$ on the authentication performance for different number of RIS elements.}
    \label{fig:N_RIS_distance}
\end{figure}
Fig. \ref{fig:N_RIS_distance} demonstrates the impact of the distance between the RIS and LT on the authentication rate for different numbers of RIS elements. As \(d_\mathrm{RL}\) increases, the authentication rate decreases across all configurations of \(N\). This behavior is primarily due to the increased path loss as \(d_\mathrm{RL}\) grows, which reduces the signal strength of the RIS-enhanced backscattered signal received at LT. 
The larger number of RIS elements enhances the beamforming capability, allowing the RIS to focus the reflected signal more effectively towards LT, thereby compensating for the increased path loss and maintaining higher authentication accuracy. 
These results also underscore the ability of RIS to extend the secure communication range in BTTNs. Although authentication performance declines with larger \(d_\mathrm{RL}\), the inclusion of RIS enables reliable authentication at distances that would otherwise result in significantly degraded performance in traditional BTTNs without RIS. As seen in Table~\ref{Table:d_TL}, acceptable authentication performance in BTTNs without RIS is limited to much shorter distances. This highlights the pivotal role of RIS in enhancing the scalability and robustness of BTTNs by enabling high-accuracy authentication even in extended distances.

\begin{figure}[t]
    \centering
    \includegraphics[width=0.36\textwidth]{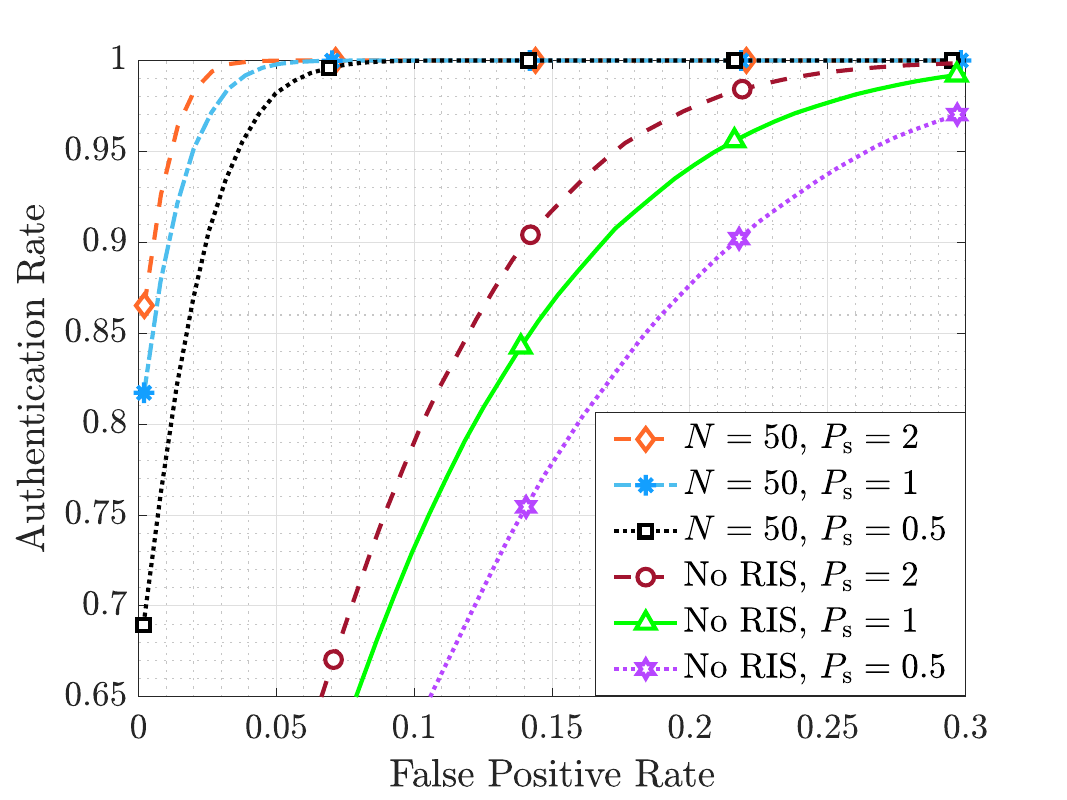} 
    \caption{ROC for different RIS scenarios and different power level of the RF source.}
    \label{fig:N_RIS_Ps}
\end{figure}
Fig. \ref{fig:N_RIS_Ps} presents the ROC curves for different RIS configurations and RF source power levels, showing their impact on authentication performance in the proposed PLA framework. The results demonstrate that higher \(P_\mathrm{s}\) improves the authentication rate across all scenarios, as stronger RF signals lead to enhanced reliability in the voltage profiles measured in LT. 
This improvement is particularly pronounced in non-RIS settings, where the lack of beamforming capabilities amplifies the reliance on source power to mitigate path loss and noise. 
On the other hand, RIS integration significantly enhances authentication performance, 
enabling higher authentication accuracy even at lower FPR values. In particular, the presence of RIS ensures robust performance even in energy-constrained scenarios, with \(P_\mathrm{s} = 0.5 \, \mathrm{dbm}\) delivering results comparable to higher power levels. This highlights RIS's capability to compensate for weaker ambient signals, making it promising for energy-efficient and secure BC.

\begin{figure}[t]
    \centering
    \includegraphics[width=0.36\textwidth]{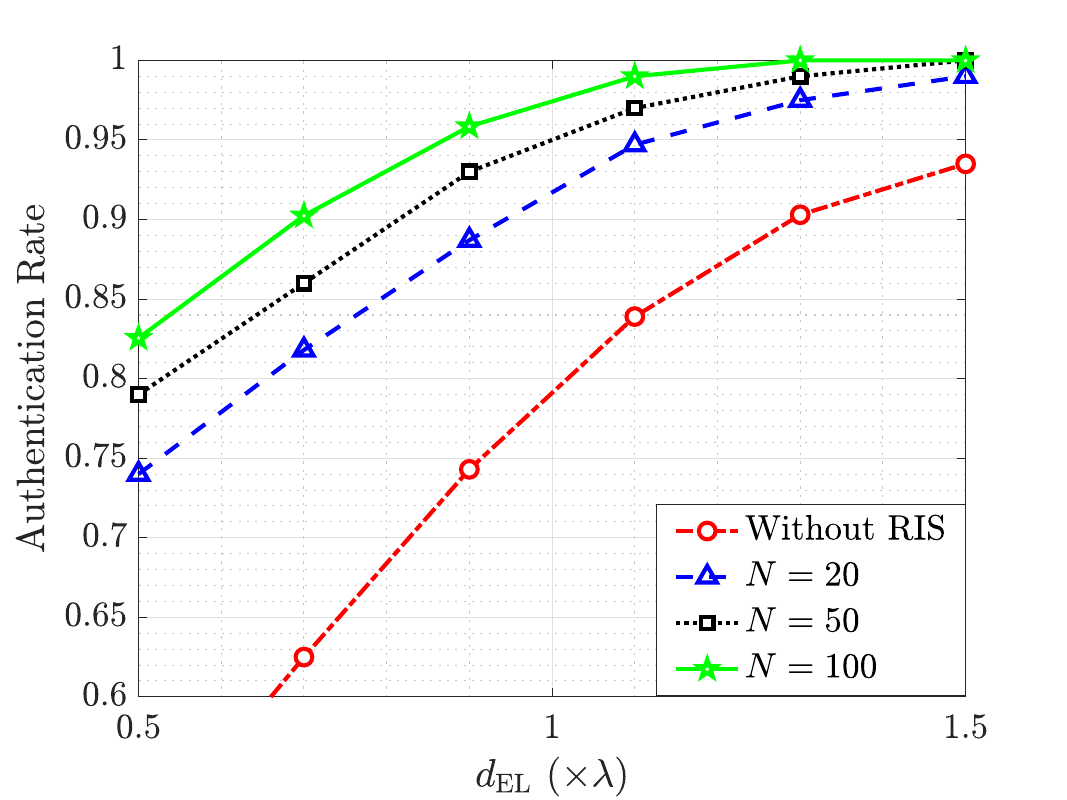} 
    \caption{Authentication performance vs. the Eve-LT distance for different RIS scenarios.}
    \label{fig:N_RIS_d_EL}
\end{figure}
Fig. \ref{fig:N_RIS_d_EL} illustrates the authentication performance as a function of the distance between Eve and LT for different RIS scenarios. It can be seen that as \(d_\mathrm{EL}\) increases, the authentication rate improves in all scenarios. This improvement is attributed to the spatial uniqueness of the CSI between TT and LT, which becomes more distinct as Eve moves farther away from LT. At larger distances, Eve's signal contributions to the received voltage profiles at LT deviate significantly from the baseline profiles, making impersonation increasingly ineffective.
The figure also highlights the RIS's ability to enhances the robustness of the authentication process even if Eve is located near to LT. 
It is important to note that Eve cannot be located closer than \(0.5 \lambda\) to LT due to physical constraints in the design of the system and the security assumptions. This limitation ensures that the CSI of the TT-LT link remains distinct and uncorrelated with the CSI of any potential Eve-LT link, thereby safeguarding the integrity and robustness of the authentication mechanism.

\begin{figure}[t]
    \centering
    \includegraphics[width=0.36\textwidth]{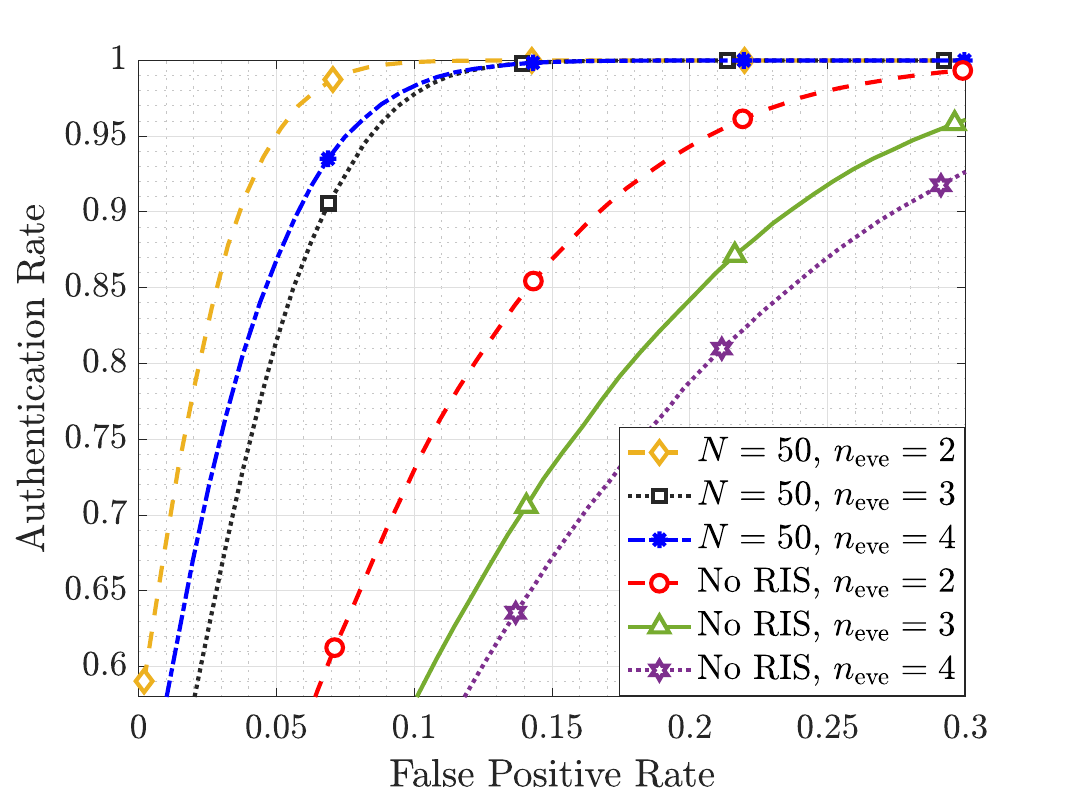} 
    \caption{ROC for the number of attackers and different RIS scenarios.}
    \label{fig:ROC_n_eve}
\end{figure}
Fig. \ref{fig:ROC_n_eve} illustrates the impact of the number of attackers (\(n_\mathrm{eve}\)) on the authentication performance under different RIS scenarios. The ROC curves depict the authentication rate as a function of FPR for \(n_\mathrm{eve} = 2, 3, 4\), both with and without RIS assistance. 
For scenarios without RIS, increasing \(n_\mathrm{eve}\) leads to a noticeable degradation in authentication robustness performance. 
Conversely, in RIS-aided cases, even as \(n_\mathrm{eve}\) increases, the authentication rate remains robust with only minimal degradation. This resilience is attributed to the RIS's ability to optimize the signal propagation and amplify the unique CSI of legitimate links, effectively mitigating the impact of multiple attackers.
Furthermore, comparing the slopes of the RIS-aided curves with those of the non-RIS scenarios reveals that the RIS-aided system exhibits significantly lower sensitivity to increases in the number of attackers, highlighting the efficacy of RIS in maintaining high authentication performance and ensuring consistent reliability, even in challenging scenarios with multiple adversaries.

\section{Conclusion 
}
\label{sec_conclusion}

This paper introduced a physical layer authentication (PLA) scheme tailored for backscattering tag-to-tag networks (BTTNs), addressing the critical challenges of secure communication in passive RFID systems. By leveraging the unique voltage profiles generated by the energy harvesting and demodulation circuits of listener tags (LT), the proposed scheme enabled lightweight and efficient authentication without imposing computational overhead. Additionally, the integration of reconfigurable intelligent surfaces (RIS) significantly enhanced the performance of the PLA by improving signal quality and extending the secure communication range, even in challenging environments. Extensive simulation results demonstrated the robustness of the proposed scheme against various attack vectors, including impersonation, replay, relay, and MITM attacks, while showcasing its ability to achieve high authentication accuracy under diverse system configurations. The findings also highlighted the transformative role of RIS in enhancing authentication performance, especially in scenarios with low ambient RF power or extended communication distances, underscoring its potential for advancing the security and scalability of future BTTNs.

\end{document}